\documentclass[12pt]{iopart}

\usepackage{iopams}
\usepackage{color}
\usepackage[utf8]{inputenc}       
\usepackage[english]{babel} 
\usepackage{amssymb,amsfonts}       
\usepackage[perpage,symbol*]{footmisc}
\usepackage[final]{graphicx}                             
\begin{document}

\title[The Quaternionic Quantum Mechanics]{The Quaternionic Quantum Mechanics}

\author{Arbab I. Arbab}

\address{Department of Physics, Faculty of Science, University of Khartoum, P.O. Box 321, Khartoum
11115, Sudan}
\ead{arbab.ibrahim@gmail.com, aiarbab@uofk.edu}

\begin{abstract}
A quaternionic wavefunction consisting of real and scalar functions is found to satisfy the quaternionic momentum eigen value equation. Each of these components are found to satisfy a generalized wave equation of the form $\frac{1}{c^2}\frac{\partial^2\psi_0}{\partial t^2}-\nabla^2\psi_0+2\left(\frac{m_0}{\hbar}\right)\frac{\partial\psi_0}{\partial t}+\left(\frac{m_0c}{\hbar}\right)^2\psi_0=0$. This reduces to the  massless Klein-Gordon equation, if we replace  $\frac{\partial}{\partial t}\rightarrow\frac{\partial}{\partial t}+\frac{m_0c^2}{\hbar}$. For a plane wave solution the angular frequency is complex and is given by $\vec{\omega}_\pm=i\frac{m_0c^2}{\hbar}\pm c\vec{k} $, where $\vec{k}$ is  the propagation constant vector. This equation is in agreement with the Einstein energy-momentum formula. The spin of the particle is obtained from the interaction of the particle with the photon field.
\end{abstract}

\maketitle
\section{Introduction}
Schrodinger equation has been generalized by Dirac and Klein-Gordon where in the former the resulting equation is a first order differential and in the latter is a second order differential equation in space and time. Dirac equation governs the motion of  spin - 1/2  particles, while Klein-Gordon governs the motion of spin - 0 particles. Dirac equation expresses the spin of the particle explicitly. Klein-Gordon equation doesn't explicitly reflect the spin of the particle. The spin of the particle shows up when a charged particle interacts with the electromagnetic field. In  Schrodinger formalism the spin of the particle is obtained by replacing the momentum of the particle by its conserved conjugate momentum in the Hamiltonian of the system. Reduction of the Dirac equation for an electron in a magnetic field to its non-relativistic limit yields the Pauli equation with a correction term which represents the interaction of the electron's intrinsic magnetic moment with the magnetic field giving the correct energy.
 In this paper we would like to address the formulation of the Klein-Gordon equation. We have recently shown the richness of describing physical laws employing \emph{Quaternions}, where we have found that the quaternionic Maxwell equations predict an existence of new scalar wave [1]. With the same fashion, we have unified the hydrodynamics and electromagnetism  in an analogous way [2]. With the same enthusiasm, we would like here to write the quantum mechanical equations governing the evolution of microscopic world in terms of Quanternions. To this end, we use the eigen value  equation, where we represent  the wave function representing the microparticle, the energy and momentum operators as quaternions. It is interesting to note that the quaternionic wave function consists of scalar and vector components. The emerging equations can then describe scalar and vector particles.  The resulting equations would then mix the scalar and vector components. We anticipate that these two particles could be the bosons and fermions. Hence, one can in some respect unify Klein-Gordon and Dirac equations in a single equation. The spin of the particle shows up when the particle interacts with the photon field. Moreover, the quaternionic uncertainty relation reduces to the ordinary uncertainty relations. This quaternionic relation yields the known Einstein's energy momentum relation.
\section{The model}
The quaternionic momentum eigenvalue problem can be written as
\begin{equation}
\widetilde{P}\,\widetilde{\Psi}=m_0c\,\widetilde{\Psi}\,
\end{equation}
where
\begin{equation}
\widetilde{P}=\left(\frac{i}{c}\,E\,, \vec{p}\right)\,,\qquad \widetilde{\Psi}=\left(\frac{i}{c}\,\psi_0\,, \vec{\psi}\right)\,.
\end{equation}
Equation (1) is in fact similar to Dirac equation [3]
 \begin{equation}
 \hat{P}\,\psi=m_0c\,\psi\,,\qquad \hat{P}=i\,\hbar \gamma^\mu\partial_\mu\,.
 \end{equation}
  Equation (1) is regarded as the quaternionic Dirac's equation.
\\
The product of the two  quaternions, $\widetilde{A}=\left(a_0\,, \vec{a}\right)$ and $\widetilde{B}=\left(b_0\,, \vec{b}\right)$ is given by [4]
\begin{equation}
\widetilde{A}\,\widetilde{B}=\left(a_0b_0-\vec{a}\cdot\vec{b}\,\,,\,\, a_0 \vec{b}+ \vec{a}\,b_0+ \vec{a}\times\vec{b}\right)\,.
\end{equation}
Apply  Eq.(2) in (1) and use Eq.(4) to get
\begin{equation}
\vec{\nabla}\cdot\vec{\psi}-\frac{1}{c^2}\frac{\partial \psi_0}{\partial t}-\frac{m_0}{\hbar}\,\psi_0=0\,,
\end{equation}
\begin{equation}
\vec{\nabla}\psi_0-\frac{\partial \vec{\psi}}{\partial t}-\frac{m_0c^2}{\hbar }\,\vec{\psi}=0\,,
\end{equation}
and
\begin{equation}
\vec{\nabla}\times\vec{\psi}=0\,,
\end{equation}
where we have used the operators: $\vec{p}=-i\hbar\vec{\nabla}$ and $E=i\hbar\frac{\partial}{\partial t}$ as in ordinary quantum mechanics.
Equations (5) -  (7) are first order linear differential equations. They can be compatible with Lorentz transformation. The scalar equation above represents the energy equation, while the vector equation represents the momentum equation. In Dirac formalism, an electron, for instance, is described by a four component wavefunction (spinors). In the present formalism the particle is described by a quaternion wavefunction having also four components. Here, one component is a scalar and the three components are vector components. We may however ask the question that: are the two formalisms equivalent to each other? \\
Equations (5) - (7) suggest that the wave functions; $\vec{\psi}$ and $\psi_0$ are not unique so that  $\vec{\psi'}$ and $\psi_0'$  can also satisfy Eqs.(5) - (6). This is possible if we define
\begin{equation}
\vec{\psi'}=\vec{\psi}+\vec{\nabla}\Lambda\,, \qquad \psi_0'=\psi_0+\frac{\partial \Lambda}{\partial t}\,,
\end{equation}
where $\Lambda$ is some scalar function. Substituting Eq.(8) in Eq.(5) yields
\begin{equation}
\nabla^2\Lambda-\frac{1}{c^2}\frac{\partial^2\Lambda}{\partial t^2}-\left(\frac{m_0}{\hbar}\right)\frac{\partial\Lambda}{\partial t}=0\,,
\end{equation}
if Eq.(5) is to be invariant under this transformation. Similarly, substituting Eq.(8) in Eq.(6) yields
\begin{equation}
\vec{\nabla}\Lambda=0\,,\qquad if\,\,\,\,\,\,\, m_0\ne 0\,,
\end{equation}
so that Eqs.(9) and (10) imply that
\begin{equation}
\frac{\partial \Lambda}{\partial t}+\frac{m_0c^2}{\hbar}\Lambda=0\,,\qquad \Rightarrow\qquad \Lambda=A\exp\,(-\frac{m_0c^2}{\hbar}\,t)\,,\qquad\rm A=const.,
\end{equation}
hence
\begin{equation}
\vec{\psi'}=\vec{\psi}\,, \qquad \psi_0'=\psi_0-\frac{m_0c^2}{\hbar}A\exp\,(-\frac{m_0c^2}{\hbar}\,t)\,.
\end{equation}
However, if $m_0=0$, then $\Lambda$ satisfies the wave equation
$$
\nabla^2\Lambda-\frac{1}{c^2}\frac{\partial^2\Lambda}{\partial t^2}=0\,.
$$
In this case the transformation in Eq.(8) is similar to the gauge transformation exhibited by the electromagnetic vector $A$ and scalar $\phi$.
Therefore, Eq.(5) - (7) are invariant under the transformation carried out in Eq.(12).

Taking the $\vec{\nabla}$ of both sides of Eq.(5) and using Eq.(6), (7)  and  the identity $\vec{\nabla}\times(\vec{\nabla}\times\vec{\psi})=\vec{\nabla}(\vec{\nabla}\cdot\vec{\psi})-\nabla^2\vec{\psi}$, one gets
\begin{equation}
\frac{1}{c^2}\frac{\partial^2\vec{\psi}}{\partial t^2}-\nabla^2\vec{\psi}+2\left(\frac{m_0}{\hbar}\right)\frac{\partial\vec{\psi}}{\partial t}+\left(\frac{m_0c}{\hbar}\right)^2\vec{\psi}=0\,.
\end{equation}
Similarly taking the divergence of Eq.(6) and using Eq.(5), one obtains
\begin{equation}
\frac{1}{c^2}\frac{\partial^2\psi_0}{\partial t^2}-\nabla^2\psi_0+2\left(\frac{m_0}{\hbar}\right)\frac{\partial\psi_0}{\partial t}+\left(\frac{m_0c}{\hbar}\right)^2\psi_0=0\,.
\end{equation}
 Alternatively, Eq.(14) can be derived if we differentiate Eq.(5) partially with respect to time  and use Eq.(6). Equation (13) can be obtained if we differentiate Eq.(6) partially with respect to the time and use Eq.(5) and the vector identity $\vec{\nabla}\times(\vec{\nabla}\times\vec{\psi})=\vec{\nabla}(\vec{\nabla}\cdot\vec{\psi})-\nabla^2\vec{\psi}$. Equations (5) and (6) represent  vector and scalar waves. To my knowledge, they are new equations. It is interesting to see from Eqs.(14) and (15) that both wavefunction components satisfy the same wave equation. Moreover, Eqs.(13) and (14) are linear equations and hence they satisfy the superposition principle of quantum mechanics. For a massless particle ($m_0=0)$ they  reduce to the ordinary wave equation, viz., $\nabla^2\psi_0-\frac{1}{c^2}\frac{\partial^2\psi_0}{\partial t^2}=0$. Notice that Eqs.(13) and (14) satisfy the transformation in Eq.(12).
They can be written in operator form as
\begin{equation}
\hat{\Pi}^2\,\vec{\psi}=0\,,\qquad \hat{\Pi}^2\,\psi_0=0\,,
\end{equation}
where the operator $\hat{\Pi}$ is defined by
\begin{equation}
\hat{\Pi}^2=\frac{1}{c^2}\frac{\partial^2}{\partial t^2}-\nabla^2+\frac{2m_0c}{\hbar\,c}\frac{\partial}{\partial t}+\frac{m_0^2c^2}{\hbar^2}\,.
\end{equation}
This can be casted in the form
$$
\hat{\Pi}^2= \frac{1}{c^2}\frac{\partial^2 }{\partial\tau^2}-\nabla^2\,,\qquad {\rm where}\qquad \frac{\partial }{\partial\tau}=\left( \frac{\partial}{\partial t}+\frac{m_0c^2}{\hbar}\right).
$$
It is thus apparent that Eqs.(13) and (14) when written in terms of the time $\tau$ become Lorentz invariant.
Equations (13) and (14) are similar to the evolution of the electric field in lossy medium [5].

If we consider a plane wave solution of the form
\begin{equation}
\psi_0=C\exp i(\omega \,t-\vec{k}\cdot \vec{r})\,,\qquad C=\rm const.,
\end{equation}
and apply it in Eqs.(13) or (14), we obtain the dispersion relation
\begin{equation}
\omega^2-2i\frac{m_0c^2}{\hbar}\, \omega-\left(\frac{m_0c^4}{\hbar^2}+c^2k^2\right)=0\,.
\end{equation}
The solution of Eq.(18) is
\begin{equation}
\omega_\pm=\frac{m_0c^2}{\hbar}\,i\pm c\,k\,.
\end{equation}
This shows clearly that $\omega$ is complex. This is a quite interesting result. It can be interpreted as having two waves moving in opposite directions (because of the term, $\pm \,c\,k$). The group velocity of these waves is the speed of light in vacuum ($v_g=\frac{\partial \omega_\pm}{\partial k}=\pm \,c$). The phase velocity is given by $v_p=\frac{\omega_\pm}{k}=\frac{m_0c^2}{\hbar\, k}\,i\pm c$. The two solutions in (19) may also represent a particle with energy $\hbar\,\omega_+$ and antiparticle with energy $\hbar\,\omega_-$. The difference of their energy states is $\Delta  E=\pm 2\hbar\, kc=\pm2\hbar\,\omega_p$, where $\omega_p$ is the photon frequency. This means that if one state is jumped to the other state two photons will be emitted (or absorbed). This interpretation resembles the Dirac theory of antiparticles.
It is an interesting  relation because it combines  the wave and particle properties. It describes a matter wave associated with particles as in de Broglie theory.

Now Eq.(19) yields
\begin{equation}
\hbar^2|\omega|^2=m^2_0c^4+k^2\hbar^2c^2\,,
\end{equation}
so that
\begin{equation}
m_0c=\hbar\,k_{\rm eff.}\,,\qquad k_{\rm eff.}=\sqrt{\frac{|\omega_\pm|^2}{c^2}-k^2}\,,
\end{equation}
is the particle's momentum in terms of angular frequency and wave number of the wave. Note that for light $p=\hbar\,k$. Equation (20) represents the interaction between particle and wave properties.
The refractive index of a medium is related to the frequency of the wave by the relation
\begin{equation}
\omega=\frac{kc}{n}\qquad\Rightarrow\qquad n_\pm=\pm\frac{ 1}{1+\frac{m_0c^2}{\hbar^2k^2}}-i\frac{\frac{m_0c}{\hbar\,k}}{1+\frac{m_0c^2}{\hbar^2k^2}}\,,
\end{equation}
where we used Eq.(19). The above equations shows that the refractive index is complex but $|n|=1$. It is a rotation of the vacuum refractive index in the complex plane by an angle
\begin{equation}
\alpha=\tan^{-1}\frac{\frac{m_0c}{\hbar\,k}}{1+\frac{m_0c^2}{\hbar^2k^2}}\,.
 \end{equation}
 The imaginary term in Eq.(22) indicates the amount of absorption loss when the  wave propagates through the space.

Applying Eq.(19) in Eq.(13) and (14), one obtains
\begin{equation}
\psi_0=C\exp (-\frac{m_0c^2}{\hbar}\,t) \exp i\,(\pm \,kct-\vec{k}\cdot \vec{r})\,,
\end{equation}
which represents a modulated wave with decaying amplitude.  The physical significance of Eqs.(13) and (14) is still under investigation. The decaying factor in Eq.(24) depends on the momentum of the particle. It is apparent that fastly moving particle decays larger than slowly moving one. Hence, the motion of the particle distorts the wave associated with it. Therefore, a matter wave is a distorting light wave. Or equivalently, when light wave passes on a moving particle the wave is distorted by the motion of the particle. However,  light wave is not distorted when encounters a stationary particle. Hence, Eqs.(13) and (14) represent the wave-particle interaction. It reflects the de Broglie hypothesis but in a rather different approach.
\section{Klein-Gordon equation (KG)}
It is the equation of motion of a quantum scalar or pseudoscalar field, a field whose quanta are spinless particles.  Any solution to the Dirac equation is automatically a solution to the Klein-Gordon equation, but the converse is not true. We would like in this section to deduce the KG from the set of equations given by (5) - (7). We seek the solution of the scalar part of the quaternionic wavefunction, viz., $\psi_0$. We anticipate that the vector part, $\vec{\psi}$, will govern the Dirac particles. Therefore,  as in Dirac's equation case,  in Eq.(14), we get
\begin{equation}
\Box^2\psi_0+\left(\frac{m_0c}{\hbar}\right)^2\psi_0=-2\left(\frac{m_0}{\hbar}\right)\frac{\partial\psi_0}{\partial t}\,,\qquad\Box^2=\frac{1}{c^2}\frac{\partial^2}{\partial t^2}-\nabla^2\, .
\end{equation}
This is a generalized Klein-Gordon equation with a time-dependent source term. The term on the right hand side of Eq.(25) can be seen as a dissipative (damping) term resulting from the inertia of the particle, or the motion of the particle in space-time. This implies that the inertia  of the particle distorts the wave nature of the particle from a pure wave. For a massless particle Eq.(25) is a pure wave and the distortion term vanishes. It is evident from Eq.(25) that the damping interaction is inevitable for all non-zero mass particles.
Now define the new time coordinate ($\tau$) as
 \begin{equation}
\frac{\partial}{\partial \tau}=\frac{\partial}{\partial t}+\frac{m_0c^2}{\hbar}\,.
 \end{equation}
Applying this transformation in Eq.(25), we get
\begin{equation}
\Box '\,^2\psi_0=0\,\,\,,\qquad{\rm where}\qquad\Box '\,^2=\frac{1}{c^2}\frac{\partial^2}{\partial \tau^2}-\nabla^2=\hat{\Pi^2}\, .
\end{equation}
Thus, in the $\tau$ coordinate the mass of the particle disappears. Or equivalently, due to the interaction (resistance) a massive particle appears to be massless. Thus, we can call the above transformation a killing mass transformation.  Hence, it is interesting that a massive particle in one coordinate system can appear massless in another coordinate system.  The electron is found to have an oscillatory motion with the extremely high frequency of $\omega_0=\frac{2m_0c^2}{\hbar}$  in addition to its normal linear motion. Schrodinger labeled that rapid oscillatory motion Zitterbewegung or "\emph{jittery motion}". Hence, the additional term in Eq.(26) may be attributed to this oscillatory nature of electrons. The oscillation is produced because of the  interference between positive and negative energy states as described in Eq.(19). Several physicists proposed the idea that the electron spin and magnetic moment as generated by a localized circulatory motion of the electron.
Using eq.(26), Eqs.(5) - (7) become
$$
\vec{\nabla}\cdot\vec{\psi}-\frac{1}{c^2}\frac{\partial \psi_0}{\partial \tau}=0\,,
$$
$$
\vec{\nabla}\psi_0-\frac{\partial \vec{\psi}}{\partial \tau}=0\,,
$$
and
$$
\vec{\nabla}\times\vec{\psi}=0\,.
$$
Now write $\psi_0(r,t)=\psi(r)\,\varphi(t)$ and  consider the  wavefunction where $\psi_0(r)=\rm const.$.  In this case Eq.(25) reduces to a damped simple harmonic oscillator with a dissipative (damping) term proportional to the mass of the particle, i.e.,
\begin{equation}
\frac{d^2\varphi}{dt^2}+\frac{2m_0c^2}{\hbar}\frac{d\varphi}{dt}+\left(\frac{m_0c^2}{\hbar}\right)^2\varphi=0\,.
\end{equation}
This has a solution of the form
\begin{equation}
\varphi(t)=B\exp\,(-\frac{m_0c^2}{\hbar}\,t)+D\,t\exp\,(-\frac{m_0c^2}{\hbar\,}\,t)\,,\qquad B \,\,\,, D\,\,\,: \rm const.,
\end{equation}
or
\begin{equation}
\varphi(t)=B\exp\,(-\frac{\omega_0}{2}\,t)+D\,t\exp\,(-\frac{\omega_0}{2\,}\,t)\,,\qquad B \,\,\,, D\,\,\,: \rm const.,
\end{equation}
The general solution of Eq.(25) as given by Eq.(24) now becomes
\begin{equation}
\psi_0=C\exp (-\frac{\omega_0}{2}\,t)\,\exp\,(\pm\, kc\,t-\vec{k}\cdot \vec{r})\,i\,.
\end{equation}
This wave depends on the rest mass of the particle and the propagation constant only. This shows that every mode with $m_0\ne 0$ damps exponentially to zero.

We remark here that  the energy equation in Eq.(20) coincides with the Einstein's relativistic energy-momentum formula, viz., $E^2=p^2c^2+m_0^2c^4$, where $\hbar\,k=p$. In this case Eq.(22) implies that
a dissipative particle wave is described by the refractive index
\begin{equation}
n_\pm=\pm\frac{1}{1+\frac{m_0^2c^2}{\hbar^2k^2}}-i\frac{\frac{m_0c}{\hbar\,k}}{1+\frac{m_0^2c^2}{\hbar^2k^2}}\,.
\end{equation}
The refractive index approaches that of vacuum as the particle's mass approaches the speed of light in vacuum, i.e., $n\rightarrow 1$ as $m_0\rightarrow 0$. Alternatively,  for the high energy  particle (\,$\hbar^2k^2>> m_0c^2$) one has $n\rightarrow 1$ as well.
\section{Interacting particles with electromagnetic radiation}
If the particle described by Eq.(2) interacts with an electromagnetic field defined  by
\begin{equation}
\widetilde{A}=\left(\frac{i}{c}\varphi\,\,, \vec{A} \right)\,,
\end{equation}
the momentum in  Eq.(2) becomes
\begin{equation}
\widetilde{P}\rightarrow \widetilde{P}+q\widetilde{A} \,\qquad\Rightarrow \qquad\vec{p}\rightarrow \vec{p}+q\vec{A}\,\qquad {\rm and}\qquad E\rightarrow E+q\,\varphi\,.
 \end{equation}
Applying Eq.(4) in Eq.(2) and using Eq.(33) yields
\begin{equation}
\vec{\nabla}\cdot\vec{\psi}-\frac{1}{c^2}\frac{\partial \psi_0}{\partial t}-\frac{m_0}{\hbar }\,\psi_0=0\,,
\end{equation}
\begin{equation}
\vec{\nabla}\psi_0-\frac{\partial \vec{\psi}}{\partial t}+\frac{qc}{\hbar}\vec{A}\times\vec{\psi}-\frac{m_0c^2}{\hbar }\,\vec{\psi}=0\,,
\end{equation}
\begin{equation}
\vec{\nabla}\times\vec{\psi}=\frac{q}{\hbar\,c}\left(\varphi \,\vec{\psi}+\vec{A}\,\psi_0\right)\,,
\end{equation}
and
\begin{equation}
\varphi\,\psi_0+c^2\vec{A}\cdot\vec{\psi}=0\,.
\end{equation}
These are two scalar equations and two vector equations. The scalar equation in Eq.(35) is the same as the scalar equation in Eq.(5). However, the vector equations are altered due to presence of the photon field. Now take the divergence of Eq.(35) to obtain
\begin{equation}
\frac{1}{c^2}\frac{\partial^2 \psi_0}{\partial t^2}-\nabla^2\psi_0+\frac{2m_0}{\hbar}\frac{\partial \psi_0}{\partial t}+\frac{qc}{\hbar}\,\vec{\psi}\cdot\vec{B}+\left(\frac{m_0^2c^2}{\hbar^2}+\frac{q^2}{\hbar^2}(A^2-\frac{\varphi}{c^2})\right)\psi_0=0,
\end{equation}
where we have used that fact that
$$
\vec{B}=\vec{\nabla}\times\vec{\psi}\,\,,\qquad \vec{\nabla}\cdot(\vec{A}\times\vec{\psi})=	 \vec{\psi}\cdot(\vec{\nabla}\times\vec{A})-\vec{A}\cdot(\vec{\nabla}\times\vec{\psi})\,.
$$
The Klein-Gordon equation for a charged particle of mass $m$ in an electromagnetic field, $A_\mu$, reads
\begin{equation}
\left(D^\mu D_\mu+\frac{m^2c^2}{\hbar^2}\right)\psi_0=0\,,\qquad {\rm where} \qquad D_\mu=\partial_\mu+\frac{q}{\hbar} A_\mu\,.
\end{equation}
The covariant derivative in Eq.(40) can be compared with the transformation in Eq.(26) if we set $q\varphi=m_0c^2$. Equation (39) can be written as
\begin{equation}
\frac{1}{c^2}\frac{\partial^2 \psi_0}{\partial t^2}-\nabla^2\psi_0+\frac{2m_0}{\hbar}\frac{\partial \psi_0}{\partial t}+\frac{qc}{\hbar}\,\vec{\psi}\cdot\vec{B}+\left(\frac{M\,c}{\hbar}\right)^2\,\psi_0=0\,,
\end{equation}
where
\begin{equation}
M^2=m_0^2+m_q^2\,\,, \qquad m_q=\frac{q}{c}\sqrt{A^2-\frac{\varphi^2}{c^2}}\,\,,
\end{equation}
is the effective mass of the interacting charged particle. The second term in Eq.(42) represents the  mass of the photon felt by the charged particle due to the particle nature of the photon.  It is interesting to see that this term  is vanishingly small. For an electron with velocity $\vec{v}$ one has $\varphi=\vec{v}\cdot\vec{A}$ so that
\begin{equation}
 m_q=q\,\frac{A}{c}\,\sqrt{1-\frac{v^2}{c^2}}\,\,.
\end{equation}
For the photon ($v=c$), $m_q=0$. This is unlike the relativistic mass where $m\rightarrow \infty$ as $v\rightarrow c$.
For a charged particle in a constant  vector field ($\vec{A}$),  Eq.(41) reads
\begin{equation}
\frac{1}{c^2}\frac{\partial^2 \psi_0}{\partial t^2}-\nabla^2\psi_0+\frac{2m_0c}{\hbar c}\frac{\partial \psi_0}{\partial t}+\left(\frac{M\,c}{\hbar}\right)^2\psi_0=0\,.
\end{equation}
Upon using  Eq.(26) the above equation reduces to
$$
\Box\,'\psi_0+\frac{m_q^2c^2}{\hbar^2}\,\psi_0=0\,.
$$
This is a Klein-Gordon equation representing a charged massless (inertially) particle having an induced mass $m_q$. Equation (42) represents the effective mass of the particle which is a function of space-time through $\vec{A}$ and $\varphi$. It is so amazing that the effective mass of such an interacting charged particle with electromagnetic field is directly proportional to its charge. The fourth term on the LHS of Eq.(39)  can also be linked to the spin of the particle, if we relate $\vec{\psi}$ to the spin $\vec{S}$ and $\psi_0$ by the relation
\begin{equation}
\vec{\psi}=\frac{1}{\hbar\,c}\vec{S}\,\psi_0\qquad\Rightarrow\qquad \frac{qc}{\hbar}\,\vec{\psi}\cdot\vec{B}=\left(\frac{q}{\hbar^2}\right)\,\vec{S}\,\cdot\vec{B}\,\psi_0\,,
\end{equation}
which corresponds to the interaction term ($H_{int.})$ of the spin of the particle with the magnetic field of the photon, when a proper field theoretic model is found for the present formalism, viz.,
\begin{equation}
H_{int.}=\left(\frac{q}{m}\right)\vec{S}\,\cdot\vec{B}=-\vec{\mu}\cdot\vec{B}\,\,,\qquad \vec{\mu}=-\frac{q}{m}\vec{S}\,.
\end{equation}
Using Eq.(45), the quaternion wave function can nw be defined as
\begin{equation}
\widetilde{\psi}=\frac{i}{c}\,\psi_0+\vec{\psi}=\exp(-\frac{iqc}{\hbar^2}\,\vec{S}\,\cdot\vec{B})\,\frac{i}{c}\,\psi_0\,,
\end{equation}
which means that the quaternionic wave function is a rotation (phase factor) of $\psi_0$. Applying Eq.(45) in Eq.(38) we get
\begin{equation}
\varphi=-\frac{c}{\hbar}\vec{A}\cdot\vec{S}\,.
\end{equation}
Now upon using Eq.(26) in Eq.(41), we obtain
\begin{equation}
\Box '\,^2 \psi_0+(\frac{q}{\hbar^2}\vec{S}\,\cdot\vec{B})\,\psi_0+\frac{m_q^2\,c^2}{\hbar^2}\,\psi_0=0\,,
\end{equation}
which represents the Klein-Gordon equation with a spin term. It is remarkable that the interacting Eqs.(35) - (38) reduce to the free Eqs.(5) - (7) when $\vec{A}=0$.

Taking the divergence of E.q(36) and using Eqs.(35), (38) and the vector identity $\vec{\nabla}\cdot(f\vec{A})=f(\vec{\nabla}\cdot\vec{A})+(\vec{\nabla}f)\cdot\vec{A}$, one gets
\begin{equation}
\left(\vec{\nabla}\varphi-\frac{\partial \vec{A}}{\partial t}\right)\cdot\vec{\psi}+\left(\vec{\nabla}\cdot\vec{A}-\frac{1}{c^2}\frac{\partial \varphi}{\partial t}\right)\psi_0=0 \,.
\end{equation}
If we replace $\vec{A}$ by -$\vec{A}$, which is equivalent to choosing the complex field $\widetilde{A}^*$ instead of $\widetilde{A}$, the above equation yields
\begin{equation}
\vec{E}\cdot\vec{\psi}=0\,,
\end{equation}
where
\begin{equation}
\vec{E}=-\left(\vec{\nabla}\varphi+\frac{\partial \vec{A}}{\partial t}\right)\,,\qquad \vec{\nabla}\cdot\vec{A}+\frac{1}{c^2}\frac{\partial \varphi}{\partial t}=0 \,.
\end{equation}
Applying Eq.(51) in Eq.(45) implies that $\vec{E}\cdot\vec{S}=0$, i.e., the direction of the electric field is perpendicular to the spin of the particle. It is understood that the spin of the particle is connected with the magnetic field. However, there is no interaction of the particle spin with the electric field. Thus, Eq.(51) can account for this fact. \emph{Interestingly, Dirac just ignored this term, calling it unphysical for a point charge to have a polarization! }
It is very interesting here to see that Eq.(51) is satisfied if the Lorentz gauge is guaranteed.  Hence, Eq.(49) is the generalized Klein-Gordon equation defining the spin of the particle. This is unlike the ordinary Klein-Gordon equation where the spin of the particle is not accounted for.
Applying Eq.(45) and (48) in Eq.(3) we obtain
$$\vec{S}\times\vec{p}=i\hbar\, q\left(\vec{A}-\frac{\vec{A}\cdot\vec{S}}{\hbar^2}\, \vec{S}\right).$$
This implies that $S^2=\hbar^2$. This can result from the combination of two spin- $1/2$ particles. This supports the idea of supersymmetry that a boson results from two interaction fermions. Hence, the spin-orbit interaction term can be described by
$$\vec{r}\cdot(\vec{S}\times\vec{p}\,)=-\vec{L}\cdot\vec{S}=i\hbar\, q\left(\vec{r}\cdot\vec{A}-\frac{\vec{A}\cdot\vec{S}}{\hbar^2}\, \vec{r}\cdot\vec{S}\right),$$
where $\vec{p}$, $\vec{S}$ and $\vec{A}$ are treated as operators. The average of the spin-orbit term vanishes. This implies that no such term can arise in a random motion of particles in an electromagnetic field. However, this can show up in an atom where it has been observed in Zeeman effect.
\section{The continuity equation}
The continuity equation governs the flow of any fluid. We would like to derive the continuity equation governing the particle flow represented Eqs.(5) - (7) and (35) - (38). Notice here that Eqs.(35) - (38) are the generalization of Eqs.(5) - (7). Multiply Eq.(35) from left  by $\psi_0$ and take the divergence of Eq.(36) with $\vec{\psi}$ and add the resulting equations to get
\begin{equation}
\vec{\nabla}\cdot\vec{J}+\frac{\partial \rho}{\partial t}=-\frac{2m_0c^2}{\hbar}\rho\,,
\end{equation}
where
\begin{equation}
\vec{J}=\psi_0\vec{\psi}\,\,\,,\qquad \rho=-\frac{1}{2}\left(\frac{\psi_0^2}{c^2}+|\vec{\psi}|^2\right)=-\frac{1}{2}\widetilde{\Psi}^*\,\,\widetilde{\Psi}\,.
\end{equation}
The RHS of Eq.(53) doesn't vanish as usual because of the dissipative effect of the  mass of the particle. However, when $m_0=0$, Eq.(53) reduces to the familiar continuity equation. It is interesting to note that the continuity equation is not affected by the presence of the fields ($\vec{A},\varphi$). This merit is also exhibited by Brans-Dicke theory where a scalar field is introduced to couple to gravity in addition to graviton [6]. Now apply Eq.(45) in (54) to get
\begin{equation}
\vec{J}=\frac{\psi_0^2}{\hbar c}\vec{S}\,\,\,,\qquad \rho=-\frac{1}{2}\frac{\psi_0^2}{c^2}\left(1+\frac{S^2}{\hbar^2}\right)\,.
\end{equation}
Since the current density is directly related to the spin of the particle, then this is a spin current density. In Eq.(55) $\rho$ represents spin density. The current density and probability spin density are related by the familiar relation of if $S=\hbar$, where we obtain $J=-\rho\, c$. This implies that spin wave travels at speed of light.
This equation can be written in the form
\begin{equation}
\vec{\nabla}\cdot\vec{J}+\frac{\partial \rho}{\partial t}=T=-\frac{2m_0c^2}{\hbar}\rho\,,
\end{equation}
where $T=-\frac{2m_0c^2}{\hbar}\rho$ is the torque density result from the two spin states. These two spin states behave like a dipole in the electric field of the photon. Hence, producing a polarization, $\vec{P}_s$ given by $T=-\vec{\nabla}\cdot\vec{P}_s$ [7].
Hence, Eq.(56) is transformed into
\begin{equation}
\frac{\partial \rho}{\partial t}+\vec{\nabla}\cdot(\vec{J}+\vec{P}_s)=0\,,
\end{equation}
which implies that the new current $\vec{J}_{\rm eff.}=\vec{J}+\vec{P}_s $ is conserved.
Alternatively, if we  define a new time coordinate by the relation
\begin{equation}
\frac{\partial}{\partial \eta}=\frac{\partial}{\partial t}+\frac{2m_0c^2}{\hbar}\,,
\end{equation}
then the continuity equation above reduces to
\begin{equation}
\frac{\partial \rho}{\partial \eta}+\vec{\nabla}\cdot\vec{J}=0\,.
\end{equation}
The factor of '2' appearing in the transformation in Eq.(58), when compared with  Eq.(26), can be attributed to the fact that the particle acquires two energy states associated with its spin when it is placed in a photon field. This allow us to deduce that the particle has two spin-state. Hence, with such setting  the continuity equation takes the normal form in the new time coordinate.
\section{Quaternionic uncertainty relation}
The quaternionic uncertainty relation can be written as
\begin{equation}
\left(\Delta\widetilde{r}\right)(\Delta\widetilde{P})\ge \frac{\hbar}{2}\,.
\end{equation}
Applying the multiplication rule in Eq.(4) yields
\begin{equation}
\Delta \vec{r}\cdot \Delta \vec{p}+\Delta t\, \Delta E\ge\frac{\hbar}{2}\,,
\end{equation}
\begin{equation}
\Delta \vec{r}\,\, \Delta E+c^2\Delta t\, \Delta \vec{p}=0\,,
\end{equation}
and
\begin{equation}
\Delta \vec{r} \times \Delta \vec{p}=0\,.
\end{equation}
Equation (62) implies that
\begin{equation}
\frac{\Delta r}{\Delta t}=c^2 \frac{\Delta p}{\Delta E}\qquad \Rightarrow\qquad v=c^2\frac{p}{E}\,,
\end{equation}
where $v=|\frac{\Delta r}{\Delta t}|$ is the velocity of the particle.
Equation (64) is the combination of the two relativistic relations; $E=mc^2$ and $p=mv$, where $m$ is the relativistic mass. It may  also define the group velocity of the wavepacket, since
\begin{equation}
v_g=\frac{\partial E}{\partial p}= \frac{\Delta p}{\Delta E}=\frac{c^2}{v}\,.
  \end{equation}
Equation (63) implies that
\begin{equation}
\Delta y \, \Delta p_z=\Delta z \, \Delta p_y\,,\qquad \Delta z \, \Delta p_x=\Delta x \, \Delta p_z\,,\qquad \Delta x \, \Delta p_y=\Delta y \, \Delta p_x\,.
\end{equation}
These equations imply that the fractional error in all components of the momenta and positions  of the particle in any direction is the same (or the momentum gradient), i.e., $\frac{\Delta p_x}{\Delta x}=\frac{\Delta p_y}{\Delta y}=\frac{\Delta p_z}{\Delta z}$. The quantity $\frac{\Delta p_x}{\Delta x}$ defines a current or mass rate. It thus implies that the mass current is conserved when a particle moves in space. It can be attributed to conservation of current or the continuity equation that can be written in the form $\vec{\nabla}\cdot\vec{p}+\frac{\partial m}{\partial t}=0$ for a particle with fixed volume.
\section{Conclusions}
We have investigated in this work  the quaternionic eigen value problem in quantum mechanics. We have written the wave function, energy and momentum of the particle as quaternionic quantities. The eigen value equation reveals that the scalar and vector parts of the wavefunction are governed by a new wave equation. This is a new equation that we wish it will describe bosons and fermions. The quaternionic uncertainty relation predicts the Einstein relativistic formula of energy and momentum. When the interaction of the particle with an electromagnetic field is introduced a spin term appeared in the equation of motion. The scalar equation doesn't change due to the interaction of the electromagnetic field with the particle. However, the vector equations are altered. We generalized the ordinary uncertainty relation to a quaternionic one. This generalization provides us with the energy momentum relation of Einstein and the remaining uncertainty relations. Hence, by adopting the quaternionic quantum mechanics, namely, Dirac equation, we arrived at a dissipative or generalized Klein-Gordon equation with a particle spin. This formalism gives rise to the generation of spin angular momentum of the particle when a photon field is introduced in the equation of motion. Further investigation is going on to explore the physics of these two waves.
\section{Acknowledgements}
\vspace*{-2pt}
This work is supported by the University of Khartoum research fund. We appreciate very much this support. Special thanks are due to F. Amin for the fruitful discussions. I would like to thank Sultan Qaboos university, Oman, for their hospitality, where  par of this work is carried out.
\section*{References}
$[1]$  Arbab, A. I.,  \emph{\tt On the analogy between the electrodynamics and hydrodynamics using quaternions},  to appear in the 14th International Conference on Modelling Fluid Flow (CMFF'09), Budapest, Hungary, 9-12 September (2009).\\
$[2]$ Arbab, A. I., and Satti., Z., \emph{Progress in Physics}, 2, 8 (2009).\\
$[3]$ Griffiths, D., \emph{Introduction to elementary particles}, John Wiley, (1987).\\
Bjorken, J. D., and  Drell, S. D., \emph{Relativistic Quantum Fields}, Mc-Graw, New York, (1965).\\
$[4]$  Tait, P. G.,  \emph{An elementary treatise on quaternions}, 2nd  ed., Cambridge  University Press (1873).\\
$[5]$ Harmuth, H. F.,  Barrett, T. W., and   Meffert, B., \emph{Modified Maxwell equations in quantum electrodynamics},
Singapore; River Edge, N.J.,  \emph{World Scientific}, (2001).\\
$[6]$ Brans, C. H., and Dicke, R. H., \emph{Mach's Principle and a Relativistic Theory of Gravitation},\emph{Physical Review} 124, 925 (1961).\\
$[7]$ Culcer, D., Sinova, J.,  Sintsyn, N. A.,  Jungwirth, T.,  MacDonald, A.H., and Niu, Q., \emph{Phy. Rev. Lett.},93, 046602, (2004).\\
\end{document}